# Grain Size Influence on Dynamics of Polar Nanoclusters in PMN-35%PT Ceramics: Broadband Dielectric and Infrared Spectra

V. Bovtun[1], S. Kamba[1*], S. Veljko[1], D. Nuzhnyy[1], J. Kroupa[1], M. Savinov[1], P. Vaněk[1], J. Petzelt[1], J. Holc[2], M. Kosec[2], H. Amorín[3] and M. Alguero[3]

[1]*Institute of Physics ASCR, v.v.i., Na Slovance 2, 18221 Prague, Czech Republic.*
*E-mail address: kamba@fzu.cz
[2]*Institute Jozef Stefan, Jamova 39, 1000 Ljubljana, Slovenia*
[3]*Instituto de Ciencia de Materiales de Madrid, CSIC, Cantoblanco, 28049 Madrid, Spain*

**Abstract**

Dielectric response $\varepsilon^*(f,T)$ and polar phonon spectra of coarse grain (grain size ~ 4 μm) and fine grain (grain size ~ 150 nm) ceramics of $PbMg_{1/3}Nb_{2/3}O_3$-35%$PbTiO_3$ were investigated at temperatures 10 - 900 K. $\varepsilon^*(f,T)$ in coarse-grain ceramics exhibits relaxor behavior at high temperatures and a sharp anomaly at the ferroelectric phase transition. The fine-grain ceramics exhibit mainly relaxor ferroelectric behavior with a smaller dielectric constant. The difference is explained by different relaxational dynamics of polar nanoclusters, which appear to be more stabilized at high temperatures in the fine-grain ceramics by pinning at grain boundaries. Below $T_C$, the growth of ferroelectric domains is suppressed in fine-grain ceramics as supported also by a second harmonic generation. On the other hand, polar phonon frequencies and their temperature dependences are almost independent of the grain size, but the selection rules for the cubic symmetry are not obeyed and all phonons are split due to a locally broken symmetry by polar nanoregions and chemical disorder. The lowest-frequency polar phonon undergoes partial softening down to ~ 0.1 THz near $T_C$ = 440 K in both ceramics, but the dielectric anomaly is caused predominantly by flipping and breathing of polar nanoclusters. Due to contribution of both the soft phonon mode and dielectric relaxations into the dielectric constant, the ferroelectric phase transition, which corresponds to the percolation threshold of the polar nanoregions into macroscopic domains, can be considered as a special case of crossover between the displacive and order-disorder type.

INTRODUCTION

In the last decade we have witnessed massive application of ferroelectric (FE) components in microelectronics. The most attractive are the applications of FE materials in multilayer capacitors (based on $BaTiO_3$), FE random access memories (mainly out of $PbZr_{1-x}Ti_xO_3$ and $SrBi_2Ti_2O_9$) and piezoelectric components (mainly based on $PbZr_{1-x}Ti_xO_3$). Unfortunately, it was shown that the FE properties (permittivity $\varepsilon'$, spontaneous polarization $P_s$, Curie temperature $T_C$ etc.) deteriorate with the reduced thickness of the FE films and with decreasing size of the ceramic grains (for the size effect see the review [1]). Understanding the size effect in FE materials became of interest of many physicists and material scientists. Up to now, many contradictory theoretical and experimental results have appeared on the critical size for ferroelectricity and about the reasons of the reduced $\varepsilon'$, $P_s$ and $T_C$. Several influences, both extrinsic and intrinsic, have been considered. They arise from strain, chemical and structural defects in the crystal lattice (e.g. oxygen vacancies), dead (low-permittivity) interfacial layers, grain boundaries in ceramics and polycrystalline films, etc.

The size effect was mostly investigated for $BaTiO_3$, since the modern multilayer capacitors with layer thicknesses below 1 μm require the $BaTiO_3$ grain size of the order of 100 nm, in which large reduction of $\varepsilon'$ was observed. It was shown that mainly extrinsic effects play the role in reduction of $\varepsilon'$ and $P_s$, as well as in smearing and shifts of the dielectric anomalies. Study of a nanosized $BaTiO_3$ single-crystal capacitor (thickness ~ 75 nm) demonstrated a dielectric anomaly as sharp as in the bulk showing that the extrinsic contributions can be eliminated.[2] Several theoretical papers estimated the critical thickness of 6, 4 or even 3 unit cells for the appearance of ferroelectricity.[3] Some papers show that the boundary conditions (electrodes, other dielectric layers on the thin film surface etc.) can help to stabilize the FE phase so that the FE phase was observed in films with thicknesses down to 3 [4] or even 1 unit cell.[5]

Concerning the influence of grain size on $\varepsilon'$, substantially lower $\varepsilon'$ is commonly observed in thin films compared to single crystals. Also many bulk ceramics exhibit reduced $\varepsilon'$ and it was shown that mainly grain boundaries of low local $\varepsilon'$ play the role in it.[6] This effect is especially remarkable in high-permittivity materials like $SrTiO_3$ (at low temperatures), where dramatic decrease of $\varepsilon'$ was observed with reducing the grain size.[7]

Huge $\varepsilon'$ exists also in bulk relaxor ferroelectrics (RFE). These materials are not necessary FEs at low temperatures, but they exhibit broad maximum of $\varepsilon'(T)$ at temperature $T_{max}$ which is strongly frequency dependent: $T_{max}$ remarkably increases and $\varepsilon'_{max}$ decreases with increasing frequency. These properties were explained by dynamics of polar nanoregions or nanoclusters (PNC), which appear typically up to ~300 K above $T_m$ ($T_m$ is $T_{max}$ at low-frequencies) at so called Burns temperature $T_B$.[8] The clusters are dynamic below $T_B$,[9] but probably become quasi-static 100-200 K below $T_B$ at so called $T^*$ temperature.[10,11,12] The size and concentration of clusters increases on cooling below $T^*$ and PNC finally freeze out at the freezing temperature $T_f$ or jumpwise increase into macroscopic FE domains below $T_C$.[13] Dielectric relaxation related to the PNC dynamics appears near $T_B$ in the THz range (overlapping with the polar phonon response), slows down and anomalously broadens (sometimes even splits) on cooling giving rise to almost uniform distribution of relaxation frequencies below $T_f$ responsible for the observed frequency-independent dielectric losses below $T_f$.[9,14,15]

Enhanced interest to the physics and technology of RFE rose after the work of Park and Shrout,[16] who discovered one order of magnitude stronger piezoelectric effect (in comparison with at that time best piezoelectric $PbZr_{1-x}Ti_xO_3$ ceramics) in the single crystal solid solution of typical RFE $PbMg_{1/3}Nb_{2/3}O_3$ (as well as in $PbZn_{1/3}Nb_{2/3}O_3$) and classical FE $PbTiO_3$. The best piezoelectric properties of $[PbMg_{1/3}Nb_{2/3}O_3]_{1-x}$-$[PbTiO_3]_x$ (PMN-PT) were observed for the PT concentration close to the morphotropic phase boundary (MPB) (x=0.33) between the tetragonal and rhombohedral phase. Giant piezoelectric response was explained by an easy rotation of polarization in the monoclinic phase close to MPB,[17,18] however the recent theoretical[19,20] and experimental[21] studies revealed no monoclinic phase near MPB, but the coexistence of rhombohedral and tetragonal nanodomains, in which the polarization can be easily rotated.[19]

Size effect in RFE and piezoelectrics was much less investigated than in classical FEs. Nevertheless, it can be expected that the same trend of miniaturization as in multilayer capacitors will occur soon in multilayer ceramic actuators and the problem of grain size influence on the dielectric and piezoelectric properties will become highly actual. Several papers concerned the grain size influence on $\varepsilon'$ and crystal structure in PMN [22] and PMN-PT.[23,24,25,26] However, all these papers refer only to dielectric response below 1 MHz. No studies of PMN and PMN-PT ceramics with different grain size were performed in the microwave (MW), THz and infrared (IR) range, not speaking of high temperatures near and above $T_B$, although the high-frequency response and its evolution with temperature is necessary for understanding the huge low-frequency $\varepsilon'$ and dynamics of the phase transitions.

In a broad spectral range the dielectric response of single crystal, ceramics and thin film was compared only for PMN.[15] Actually, 2-3 times lower $\varepsilon'_m$ ($\varepsilon'_m$ is $\varepsilon'_{max}$ at the lowest measured frequencies) in PMN ceramics and almost 10 times lower $\varepsilon'_m$ in thin films compared to PMN single crystal was observed and explained by different dynamics of PNC in confined ceramic grains and thin films, which leads to higher relaxation frequency and lower dielectric strength. Also a FE soft mode (SM) doublet was observed in the THz and IR spectra whose $A_1$ component partially softened on heating to $T_B$ and the E component appeared quite soft, below 1 THz at all temperatures below $T_B$. SM is split due to the locally broken non-cubic symmetry in PNC, although the macroscopic symmetry remains cubic at all temperatures. The frequencies of both SM components did not differ among all the samples, i.e. the phonon dynamics is not sensitive to the size of the sample or grain, only the dielectric relaxation (PNC dynamics) is strongly affected by the available dimensions.

The aim of this paper is the study of broad-band ($10^2$-$10^{14}$ Hz) dielectric spectra of PMN-PT (x = 0.35) ceramics, i.e. of the composition near MPB with the giant piezoelectric response. PMN-PT (x = 0.35) exhibits two successive phase transitions on cooling, from cubic to tetragonal and rhombohedral phase, but also a strong relaxor behavior in the paraelectric phase. The question arises about the influence of the fine grain structure 1) on the relaxor dielectric response, 2) on the dynamics of PNC and 3) on both FE phase transitions in PMN-PT. Therefore the dielectric properties of the coarse grain (~4 μm) and fine grain (~150 nm) ceramics will be compared in a broad frequency and temperature range. Dielectric data will be accomplished by a second harmonic generation (SHG) experiment which is sensitive also to local breaking of the centre of symmetry expected in PNC below $T_B$.

EXPERIMENTAL

Two kinds of the PMN-PT (x = 0.35) ceramics with different grain structure were prepared and studied. Fine grain ceramics (FGC) with the average grain size ~ 150 nm and ~ 97% theoretical density were prepared from the nanocrystalline powder by hot pressing.[27] Coarse grain ceramics (CGC) with the average grain size ~4 μm and ~ 92% theor. density were prepared from the same nanocrystalline powder by sintering in a PbO atmosphere.[27]

Complex dielectric response $\varepsilon^* = \varepsilon' - i\varepsilon''$ of both ceramics was studied between 5 and 900 K in the frequency range from 100 Hz to 100 THz. Several experimental techniques were used to cover such a broad spectral range. The low-frequency dielectric properties were measured in the 100 Hz – 1 MHz range using a HP 4192A impedance analyzer with a Leybold He-flow cryostat (operating range 5–300 K) and a custom-made furnace (300–900 K). Dielectric measurements in the high-frequency range (1 MHz – 1.8 GHz) were performed using a computer-controlled dielectric spectrometer equipped with a HP 4291B impedance analyzer, Novocontrol BDS 2100 coaxial sample cell and Sigma System M18 temperature chamber (operating range 100–500 K). The dielectric parameters were calculated taking into account the electromagnetic field distribution in the sample. Microwave (MW) dielectric properties were measured at 8.8 GHz using the $TE_{0n1}$ composite dielectric resonator,[28] vector network analyzer Agilent E8364B and Sigma System M18 temperature chamber (100–380 K). Time-domain THz transmission spectra were obtained using a laboratory-made spectrometer based on an amplified Ti-sapphire femtosecond laser system. Two ZnTe crystal plates were used to generate (by optic rectification) and to detect (by electro-optic sampling) the THz pulses. Both the transmitted field amplitude and phase shift were simultaneously measured which allowed us to determine directly the complex dielectric response $\varepsilon^*(\omega)$ in the range from 3 to 35 $cm^{-1}$ (100 GHz – 1 THz) at temperatures from 5 K to 800 K. An Optistat (Oxford Inst.) continuous flow cryostat with mylar windows was used for cooling, and a commercial high temperature cell SPECAC P/N 5850 for the heating. The samples are highly absorbing in the THz range, therefore thin plates with thickness of 43 μm (diameter of 8 mm) were investigated. IR reflectivity spectra were obtained using the Fourier transform IR spectrometer Bruker IFS 113v in the frequency range of 20−3300 $cm^{-1}$ (0.6–100 THz) at temperatures 5-800 K. The same furnace and cryostat as for the THz spectrometer were used, only with polyethylene windows instead of mylar ones for the cryostat.

Temperature dependences of the dielectric parameters at low, high and MW frequencies were measured during slow cooling with a temperature rate 1÷2 K/min, if not extra specified. THz and IR spectra were taken at the fixed stabilized temperatures with interval of 25-50 K.

Differential scanning calorimetry (DSC) measurements were performed on Perkin-Elmer Pyris-Diamond DSC calorimeter in the temperature range 100–600 K with a temperature rate of 10 K/min.

Second harmonic generation (SHG) was studied in the reflection geometry in the temperature range 300-750 K. Q-switched Nd-YAG laser served as a light source, SHG signal at 532 nm was detected using a photomultiplier followed by a boxcar integrator. Samples were placed in an oven allowing continuous temperature change (2-3 K/min) from 300 up to 750 K.

RESULTS AND ANALYSIS

$\varepsilon'(T)$ of CGC (see Figure 1) shows a sharp maximum corresponding to the FE phase transition at $T_C$ = 440 K and an additional anomaly at $T_{TR}$ = 270÷290 K revealing the tetragonal – rhombohedral phase transition. Temperature hysteresis (~ 5 K) of $\varepsilon'(T)$ observed between zero-field heating and cooling as well as the endothermic DSC peak on heating (not shown) indicate the first-order FE phase transition at $T_C$. On the other hand, dielectric relaxation, revealed near and above $T_C$ in the 1 MHz – 1 THz range, is characteristic of RFE.[9] Dielectric dispersion in the FE phase below $T_{TR}$ is less pronounced. The data in Figure 1 obtained with CGC are qualitatively the same as the data obtained from a single crystal,[29] only the value of $\varepsilon'$ is slightly lower, apparently due to the 8% porosity of our ceramics.

$\varepsilon'(T)$ of the FGC (see Figure 2) shows no sharp anomalies which could be attributed to phase transitions, but only typical response of RFE is seen, i.e. a diffuse $\varepsilon'(T)$ maximum at $T_m$ ≈ 430 K and a broadband relaxational dielectric dispersion both below and above $T_m$. No DSC anomaly was observed in FGC.

Low-frequency maxima of $\varepsilon'(T)$ in both CGC and FGC undergo strong dielectric dispersion of relaxational character. Below we will determine the contributions of phonons and dielectric relaxations to the whole dielectric dispersion.

Let us compare the dielectric properties of CGC and FGC. Low-frequency permittivity (100 Hz, Figure 3a) of FGC at $T_m$ is about three times smaller than that of CGC at $T_C$, while the high-frequency permittivity (300 MHz, Figure 3c) is only by 25% smaller. The maximal loss values of FGC are much smaller than those of CGC both at low and high frequencies (Figure 3b, 3d). The values and shape of the THz $\varepsilon'(T)$ and $\varepsilon''(T)$ curves in both the samples are very similar (250 GHz, Figures 3e,f), only their peaks are slightly shifted to lower temperatures in FGC. Below $T_{TR}$, i.e. at temperatures corresponding to the rhombohedral phase in CGC, there is no essential difference between CGC and FGC at any frequency. The features mentioned above show similarity of the high-frequency polarization mechanisms in both ceramics, but more pronounced low-frequency polarization mechanisms and stronger contributions of the PNC dynamics to the permittivity in CGC. The main dielectric dispersion in both ceramics takes place in the MW range, similar to PZN-8%PT single crystals.[30]

Broadband dielectric spectra of CGC (Figure 4) reveal two relaxation regions R1 and R2 between 1 MHz and 1 THz. These relaxations partially slow down on cooling, but do not shift below 1 MHz, unlike in PMN and other RFEs without phase transitions.[9,14,15] The relaxations are weak and very broad below $T_{TR}$ and nearly frequency-independent dielectric losses are observed below 200 K. Presence of two relaxational contributions (R1 and R2) in the dielectric response of CGC at $T > T_{TR}$ is clearly evidenced from the observed peaks of $\varepsilon''(f)$ and from the multi-component Cole-Cole fits (dash lines at $T \geq 300$ K in Figure 4). For the detailed analysis we used the empirical Cole-Cole equation with three relaxational contributions:

$$\varepsilon^*(f) = \varepsilon'(f) - i\varepsilon''(f) = \varepsilon_\infty + \sum_{j=1}^{3} \frac{\Delta\varepsilon_j}{1 + (if/\nu_{Rj})^{1-\alpha_j}}, \quad (1)$$

where $\Delta\varepsilon_j$ is the contribution of the relaxation to the static permittivity (dielectric strength), $\varepsilon_\infty$ is the contribution of the phonon modes and higher–frequency electronic processes to permittivity, $f$ is the frequency and $\nu_{Rj}$ is the mean relaxation frequency of the j-th relaxation. Parameter $0 \leq \alpha_j \leq 1$ characterizes the distribution of relaxation times. The lowest-frequency relaxation R3 is present only at high temperatures and low frequencies below 10 kHz. It could be caused by oxygen vacancies, but some authors interpreted it as due to a dynamics of PNC.[31] Its mean relaxation frequency $\nu_{R3}$ lies below our frequency range and was evaluated in details by other authors.[31] In the subsequent discussion below we will concentrate on evaluation of the two higher-frequency (above 1 MHz) relaxation processes only, describing the PNC dynamics to our mind.

The dielectric spectra of FGC (Figure 5) reveal only one dielectric relaxation (or peak in $\varepsilon''(f)$) in the THz - MW region, which partially slows down and broadens on cooling. Below 100 K the relaxation is weak and nearly frequency-independent dielectric losses are observed. Two relaxational contributions R1 and R2 clearly seen in CGC cannot be unambiguously determined from the dielectric spectra of FGC, but they are recognizable via the asymmetric shape of the $\varepsilon''(f)$ peaks marked as R in Figure 5b. The relaxation appears mainly above 10 GHz (i.e. at higher frequencies than in CGC) which is responsible for the lower dielectric strength of this relaxation and the resulting lower $\varepsilon'$ of FGC at low frequencies.

For the detailed analysis of the temperature evolution of relaxation frequencies, also the THz and IR spectra are needed, because the relaxation shifts to THz region at high temperatures. THz spectra above 200 GHz are shown in Figures 4 and 5, the IR reflectivity spectra in Figure 6. The reflectivities $R(\omega)$ below 1 THz were calculated from the complex THz permittivity spectra $\varepsilon^*(\omega)$ using the formula

$$R(\omega) = \left| \frac{\sqrt{\varepsilon^*(\omega)} - 1}{\sqrt{\varepsilon^*(\omega)} + 1} \right|^2. \quad (2)$$

The shape of the IR spectra is very similar to the spectra of pure PMN [32] and PMN-27%PT [33] crystals or PZN-PMN-PSN ceramics.[34] The IR spectra of CGC and FGC are similar, only the reflectivities of CGC (Figure 6a) are slightly lower apparently because of its lower density. It causes higher diffuse scattering of the reflected IR beam (rising with frequency) and a lower reflectance above 200 cm$^{-1}$ in the less dense ceramics.

IR reflectivity and THz dielectric spectra were fitted simultaneously using Eq. (2) and a generalized oscillator model with the factorized form of the complex permittivity

$$\varepsilon^*(\omega) = \varepsilon_\infty \prod_{j=1}^{n} \frac{\omega_{LOj}^2 - \omega^2 + i\omega\gamma_{LOj}}{\omega_{TOj}^2 - \omega^2 + i\omega\gamma_{TOj}}. \tag{3}$$

Here $\omega_{TOj}$ and $\omega_{LOj}$ mark the transverse and longitudinal frequencies of the j-th mode, respectively, and $\gamma_{TOj}$ and $\gamma_{LOj}$ denote their corresponding damping constants. The high-frequency permittivity $\varepsilon_\infty$ has its origin in electronic absorption processes at much higher frequencies than phonon frequencies (typically in UV-VIS range) and it was obtained from the frequency-independent room-temperature reflectivity above the phonon frequencies. The temperature dependence of $\varepsilon_\infty$ is usually very small and was neglected in our fits.

The real and imaginary parts of $\varepsilon^*(\omega)$ obtained from the fits of IR and THz spectra of CGC and FGC are shown together with the experimental THz data in Figures 7÷10. $\varepsilon'$ and $\varepsilon''$ values increase in the THz range on heating up to 450 K (close to $T_C$), but on further heating they decrease. Therefore for clarity we plotted the $\varepsilon^*$ spectra of each sample in two figures (below and above 450 K).

Three main reflection bands characteristic for the simple cubic paraelectric perovskite structure (TO1, TO2 and TO4 modes (TO3 is silent)) with $Pm\bar{3}m$ space group are seen in the IR reflectivity (Figure 6) as well as in $\varepsilon''(\omega)$ spectra of both ceramics (Figures 7-10). Actually 7 oscillators were necessary for the fit of paraelectric phase at 800 K. Such a number of IR active phonons can be expected from the factor group analysis of neither $Pm\bar{3}m$ nor $Fm\bar{3}m$ paraelectric phase, where only 3 and 4 polar modes are allowed, respectively. One could assume splitting of each mode in the $Fm\bar{3}m$ structure due to chemical inhomogeneities in the perovskite B-sublattices (for details see [35]). 13 modes were used for the fit of IR spectra in FE phase at 5 K, which corresponds to the rhombohedral structure with 1:1 short range order in the B perovskite site, where 16 IR active modes are allowed.[35] The phonon spectra and their evolution with temperature are very similar in both kinds of ceramics (see e.g. the modes below 100 cm$^{-1}$ in Figure 11).

The lowest-frequency reflection band seen below 100 cm$^{-1}$ is due to the TO1 phonon which plays the role of the FE soft mode (SM) and describes mainly the vibration of Pb-cations against the BO$_6$ octahedra framework (so-called Last mode[36]). This band is actually split into two components of A$_1$(TO1) and E(TO1) symmetry (see Figure 11) not only below $T_C$ due to the rhombohedral symmetry, but also due to the local uniaxial symmetry in PNC above $T_C$. The E(TO1) component is responsible for the increase in the reflectivity below 50 cm$^{-1}$ and it is also very well pronounced in the $\varepsilon''(\omega)$ spectra in Figures 7 and 9.

A$_1$(TO1) mode is underdamped ($\gamma_{TO} < 2\omega_{TO}$) below and overdamped ($\gamma_{TO} > 2\omega_{TO}$) above ~ 450 K in both ceramics. On lowering temperature the A$_1$(TO1) frequency increases from 52 cm$^{-1}$ and saturates near 81 cm$^{-1}$ below 250 K. Below the same temperature two new modes appear near 78 and 50 cm$^{-1}$. They presumably stem from the Raman active mode of F$_{2g}$ symmetry in the paraelectric phase, which activate in the FE phase below 250 K.

It is interesting to note that the A$_1$(TO1) phonon was recently investigated in PMN-PT single crystal by means of inelastic neutron scattering (INS).[37] The authors found that on cooling the phonon hardens only below $T_C$ while we see the hardening already from $T_B$. This is because the so-called phonon waterfall occurs in the INS spectra due to the SM coupling with the transverse acoustic phonon branch.[38] The TO1 phonon frequency cannot be directly determined from INS at the Brillouin-zone center ($q = 0$), its frequency is estimated only from the extrapolation of the optical phonon branch for $q \neq 0$.[37] On the other hand, IR spectroscopy can see only the long-wavelength polar phonons with $q \approx 0$ with higher resolution and accuracy than INS.

The E(TO1) mode frequency behaves with temperature change similarly in both kinds of ceramics (Figure 11) and exhibits minimum ($\omega_E$ ~14 cm$^{-1}$) near 450 K, i.e. near $T_C$ or $T_m$. Therefore it reminds the FE soft mode, although the phase transition is not distinct in the dielectric data of FGC (Figure 2).

The E(TO1) SM is mostly overdamped and, moreover, above room temperature it is overlapping with the relaxation mode (Figure 12) describing the dynamics of PNC. While the A$_1$(TO1), TO2 and TO3 phonon contributions to $\varepsilon'$ are moderate, the E(TO1) contribution is large and achieves (according to the IR and THz fits) more than 2000 between 450 K and 600 K. Note that the static $\varepsilon'$ is still several times higher due to the contribution of the relaxational modes seen if Figures 4 and 5. The relaxation frequencies increase on heating, approach the THz range and merge with the E(TO1) component at temperatures above $T_B \approx 600$ K. This is well known also from the pure PMN [15] and other RFEs.[38] Therefore the overdamped feature seen above 300 K in the THz spectra below 1 THz should be assigned to E mode overlapped with the relaxation. The relaxation is frequently called central mode since in inelastic scattering experiments it appears as a central component of the spectra.

The fit parameters of an overdamped oscillator (Eq. 3) are not unambiguous, because the eigenfrequency $\omega_{SM}$ and damping $\gamma_{SM}$ (hereafter we will speak only about the E component of the SM) are correlated and better physical meaning has the ratio

$$\omega_{SM}^2/\gamma_{SM} = \nu_{SM}, \tag{4}$$

which very approximately corresponds to the maximum of dielectric loss $\varepsilon''(\omega)$. We renormalized $\omega_{SM}$ seen in Figure 11 using Eq. (4), transformed the frequency from cm$^{-1}$ to Hz and plotted it in Figure 12 together with the mean relaxation frequencies R1, R2 and R corresponding to the experimental points of $\varepsilon''(f)$ maxima seen in Figures 4 and 5.

Dielectric spectra of CGC clearly reveal two relaxation modes R1 and R2. Frequency of the high-frequency mode ($\nu_{R1}$) increases quickly with temperature and arrives to the THz range near 300 K (see Figures 7-10 and 12), while the low-frequency mode frequency ($\nu_{R2}$) remains below 1 GHz at least till 500 K. Temperature dependence of $\nu_{R2}(T)$ does not follow a simple exponential law, but shows an anomaly near $T_C$ = 440 K, which resembles a critical slowing down for the order-disorder ferroelectrics.[39,40] In addition, a quick exponential-like decrease of $\nu_{R2}$ appears on cooling in the FE phase below 350 K.

To describe the $\nu_{R2}(T)$, we combine the Vogel-Fulcher law usually used in the case of RFEs

$$\nu = \nu_0 \cdot \exp[-U/(T-T_{VF})] \tag{5}$$

and the critical slowing down[39,40]:

$$\nu = A_1 \cdot (T_C - T), \quad \text{at} \quad T < T_C, \tag{6a}$$

$$\nu = A_2 \cdot (T - T_C), \quad \text{at} \quad T > T_C, \tag{6b}$$

where $\nu_0$ is a high temperature limit of the relaxation frequency, $U$ and $T_{VF}$ are the Vogel-Fulcher activation energy and temperature, respectively, and $A_1$ and $A_2$ are temperature independent constants.

Accounting that the phase transition in CGC is of the first order, we derived the following equations with different parameters below and above $T_C$:

$$\nu_{R2} = A_1 \cdot (T_1 - T) \cdot \exp[-U_1/(T-T_{VF1})], \quad \text{at} \quad T < T_C, \tag{7a}$$

$$\nu_{R2} = A_2 \cdot (T - T_2) \cdot \exp[-U_2/(T-T_{VF2})], \quad \text{at} \quad T > T_C, \tag{7b}$$

where $T_1$ and $T_2$ or $T_{VF1}$ and $T_{VF2}$ are the critical temperatures and Vogel-Fulcher temperatures, respectively, and $U_1$ and $U_2$ are the Vogel-Fulcher activation energies below and above $T_C$, respectively. Solid line R2 in Figure 12 corresponds to the fit of $v_{R2}(T)$ with the Equations (7a,b) using parameters shown in Table 2. One can see that combination of the critical slowing down and Vogel-Fulcher law describes well the experimental temperature dependence $v_{R2}(T)$. The critical temperatures $T_1$ and $T_2$, obtained from the fit, are close to $T_C$. The values of the Vogel-Fulcher temperature and activation energy are typical for PMN and other Pb-containing RFEs.[14,15]

Temperature dependence of $v_{R1}(T)$ generally follows the Arrhenius law, but also shows an anomaly near $T_C$ (see curve R1 at Figure 12). To describe the $v_{R1}(T)$, we combine the Arrhenius law

$$v = v_0 \cdot \exp[-E/T] \qquad (8)$$

with the critical slowing down (6) and fit $v_{R1}(T)$ with the following equations:

$$v_{R1} = A_1 \cdot (T_1 - T) \cdot \exp[-E_1/T], \quad \text{at} \quad T < T_C, \qquad (9a)$$

$$v_{R1} = A_2 \cdot (T - T_2) \cdot \exp[-E_2/T], \quad \text{at} \quad T > T_C, \qquad (9b)$$

where $v_0$ is a high temperature limit of the relaxation frequency, $E$, $E_1$, $E_2$ are the Arrhenius activation energies. The fit parameters are shown in Table 2.

The observed combinations of the critical slowing down with the Arrhenius or Vogel-Fulcher temperature dependence of $v_{R1}$ and $v_{R2}$ reflect the coexistence of the FE transition and relaxor behavior in CGC. The facts that activation energies are the same below and above $T_C$ ($U_1 = U_2 = 700$ K, $E_1 = E_2 = 1500$ K) and Vogel-Fulcher temperatures are very similar ($T_{VF1}$, $T_{VF2} \sim 200$ K), show that the relaxor behavior takes place not only above but also below $T_C$. Similar to PMN,[15] R1 and R2 can be attributed to the dynamics of polar nanoclusters, i.e. to their breathing and flipping. In the case of PMN and PMT [15,35] it was shown that the relaxation describing flipping follows the Vogel-Fulcher law, while the breathing follows the Arrhenius law. The FE transition in PMN-PT can be consequently considered as a result of a stepwise increase in the size of the nanoclusters, i.e. percolation of PNC into macroscopic domains below $T_C$. Such interpretation explains why the relaxations slow down only partially on cooling unlike in PMN and other RFE without macroscopic structural phase transitions.[9,13,14,15] In our samples the PNC dynamics contributes to the dielectric response even below $T_C$ which means that obviously not all PNC have transformed into FE domains. Our result is supported by the piezoresponse force microscopy of PMN-20%PT single crystals, which actually observed "normal" FE domains into which nanodomains of opposite $P_s$ orientation are embedded.[41]

Two-component relaxation was observed also in the Brillouin scattering study in the GHz and THz range.[42] The authors explained it by two types of dipole moment rotation in PNC. The fast relaxation was attributed to the 180° dipole flipping and the other one to the non-180° dipole flipping. In our opinion, the flipping is not probable at low temperatures in the FE phase, only the breathing (i.e. fluctuations of the PNC volume or in other words PNC-wall motion) can be active in analogy to the well-known FE domain-wall dynamics. Therefore we see only one broad relaxation below 200 K due to the PNC breathing.

Coexistence of the FE domains and PNC in CGC below $T_C$ ($T_m$) is also supported by the SHG experiment (Figure 13). The SHG signal in CGC does not tend to saturate on cooling below $T_C$ unlike in classical ferroelectrics,[43] but increases linearly down to room temperature, probably due to additional contribution of PNC, whose concentration or volume changes below $T_C$. The SHG signal is relatively large in the FE phase and its sharp decrease on heating above ~ 440 K can be attributed to the first order FE transition. The weak tail above $T_C$ which disappears only near $T_B$ ~ 650 K gives evidence for the presence of PNC in the cubic phase. Thus, the temperature evolution of the dielectric spectra as well as SHG signal indicates presence of polar nanoclusters in CGC both below and above $T_C$.

Temperature dependence of SHG signal in FGC (Figure 13) shows no step-like changes, but only a smooth gradual change of the slope near $T_C$ = 440 K. It is important to stress that the SHG signal in FGC above $T_C$ is much higher than in CGC. It gives evidence about a higher concentration of PNC in macroscopically cubic structure of FGC. Moreover, the SHG signal remains nonzero up to our highest investigated temperature 750 K (Figure 13). It means that the Burns temperature in FGC is markedly higher than that in CGC. Enhanced stability of PNC in FGC can be related to their reduced ability to flip under the applied electric field probably due to pinning at the grain boundaries. Below $T_C$ the SHG signal is much smaller than in CGC, giving evidence about partial suppression of the FE domain growth. Nevertheless, the SHG intensity remarkably increases on cooling below $T_C$ so that some smeared diffuse FE transition probably still occurs in FGC. This is fully in agreement with a previous work on analogous samples that showed submicron/nanometer sized crosshatched domains for the FGC sample and micron-sized lamellar domains for the CGC.[25]

Relaxational central mode component R in FGC can be also attributed to the dynamics of PNC. Its mean frequency $\nu_R$ is related to superposition of both flipping and breathing mechanisms, which cannot be unambiguously separated from the spectra. This appears also due to a much smaller contribution of the PNC flipping, which is consequence of the above mentioned pinning of PNC on the grain boundaries. Temperature dependence of the relaxation frequency $v_R(T)$ (Figure 12) follows roughly the Arrhenius law (8) typical for a temperature activated process. The parameters of the fit are shown in Table II. No anomalies of $v_R(T)$ were observed near $T_C$. Nevertheless, similar to CGC, the relaxation does not shift below 1 MHz which may be due to percolation of some of the PNC into FE domains below $T_C$. It is seen that cluster flipping strongly contributes to the permittivity in RFE crystals [15] and coarse grain ceramics, however in the fine grain ceramics, where the probability of PNC flipping is reduced, the permittivity is much lower (see Figures 1 and 2).

We attempted to explain the difference in the dielectric dispersion of FGC and CGC (Figures 4 and 5) by considering a reduced permittivity at the grain boundaries (so called dead layers). The grain-size reduction results in increase of the relative dead-layer volume and the following reduction of the effective $\varepsilon'$. Such approach was successfully used to explain the suppressed effective $\varepsilon'$ observed in SrTiO$_3$ ceramics and thin films.[6,7] We performed simulations according to the coated spheres (core-shell) model as in Refs. [44,45]. If we consider the dielectric spectra of CGC as a bulk property of cores and a small frequency-independent $\varepsilon'$ as the property of the shells, the effective dielectric spectra of the core-shell composite still consisted of two well-separated relaxations. But the experimental spectra of the FGC show only one smeared relaxation which is closer to R1 relaxation in CGC. While some effect of the lower $\varepsilon'$ in the grain boundaries still cannot be excluded, it clearly shows that the main effect is, however, suppression of the R2 relaxation connected with the PNC flipping due to their pinning by the grain boundaries. Pinning of the PNC and blocking of the FE domain growth below $T_C$ seems to be responsible for the domination of the relaxor behavior (Figure 2) in FGC.

CONCLUSIONS

In spite of some differences and peculiarities, coarse and fine grain PMN-35%PT ceramics have a common feature: similar phonon dynamics, existence of the relaxational dynamics of PNC, broad-band dielectric response and temperature evolution of the local polarization demonstrated from the SHG experiment and characterized by a superposition of the relaxor behavior and the FE transition. In CGC, clear coexistence of the first order FE transition near 440 K and relaxor behavior with the Burns temperature of ~ 650 K was revealed. Reduction of the grain size in FGC resulted in weakening and smearing of the phase transition features and domination of the relaxor behavior with higher estimated Burns temperature above 750 K.

The lowest-frequency TO1 polar phonon is split for all temperatures into the E and $A_1$ components due to a local dielectric anisotropy in PNC. The $A_1$ component hardens on cooling and saturates below $T_{TR}$. The E component lies below 1 THz at all temperatures, and decreases down to ~ 0.1 THz near $T_C$. Nevertheless, this softening is not responsible for the whole dielectric anomaly observed near $T_C$ (or $T_m$), although it is apparently the driving force of the phase transition. The high value of the low-frequency permittivity is mainly caused by microwave dielectric relaxation(s). The relaxation is split into two components in CGC. The lower-frequency component R2 is assigned to flipping of PNC while the higher-frequency component R1 to their breathing. The relaxation frequency $\nu_{R2}$ of the flipping follows the Vogel-Fulcher law combined with partial critical slowing down near $T_C$ and the relaxation frequency $\nu_{R1}$ of the breathing follows the Arrhenius law combined with a small critical slowing down near $T_C$. Both relaxations are seen not only above $T_C$ but also in the FE phase. It gives evidence, together with the temperature dependence of SHG signal, about the coexistence of FE domains and PNC in the FE phase. In FGC the relaxation frequencies are overlapping and the lower-frequency relaxation R2 is highly suppressed. The effective relaxation frequency $\nu_R$ follows the Arrhenius law without any anomaly near $T_C$ and the breathing of PNC dominates in the dielectric response. It is presumably caused by pinning of PNC at the grain boundaries. Stabilization of PNC supported by the SHG experiment up to high temperatures indicates higher Burns temperature in the FGC. On other hand, smaller SHG signal compared with that in CGC below $T_C$ indicates dominance of the PNC over the normal FE domains.

ACKNOWLEGEMENTS

The authors acknowledge the support of the Czech Science Foundation (project No. 202/06/0403), of the Academy of Sciences of the Czech Republic (projects AVOZ 10100520 and KJB 100100704) and of the Ministry of Education (COST-OC 101).

FIGURE CAPTIONS

FIG. 1. (Color online) Temperature dependences of (a) the dielectric permittivity $\varepsilon'$ and (b) loss $\varepsilon''$ of the CGC for frequencies between 100 Hz and 250 GHz.

FIG. 2. (Color online) Temperature dependences of the dielectric permittivity (a) $\varepsilon'$ and (b) loss $\varepsilon''$ of the FGC at frequencies between 100 Hz and 250 GHz.

FIG. 3. (Color online) Temperature dependences of the dielectric permittivity $\varepsilon'$ (a, c, e) and loss $\varepsilon''$ (b, d, f) of the CGC (solid lines) and FGC (dash lines) at 100 Hz (a, b), 300 MHz (c, d) and 250 GHz (e, f).

FIG. 4. (Color online) Frequency dependences of (a) dielectric permittivity $\varepsilon'$ and (b) loss $\varepsilon''$ of the CGC at temperatures between 100 K and 480 K. Dash lines at T ≥ 300 K are multi-component Cole-Cole fits. R1 and R2 denote loss maxima of the two relaxational contributions (see the text).

FIG. 5. (Color online) Frequency dependences of (a) dielectric permittivity $\varepsilon'$ and (b) loss $\varepsilon''$ of the FGC at temperatures between 100 K and 450 K. R denotes the asymmetric loss maximum by overlapping two relaxational contributions (see the text).

FIG. 6. (Color online) IR and THz reflectivity spectra of (a) CGC and (b) FGC at temperatures between 5 K and 800 K. THz reflectivities were calculated from the THz dielectric response.

FIG. 7. (Color online) Complex dielectric response of the CGC in the THz and IR range (solid lines) obtained by fitting of the IR reflectivity spectra together with THz experimental data (symbols) at temperatures below 450 K. TO1, TO2 and TO4 denote the loss maxima attributed to the 3 IR-active transverse optical phonons. $A_1$(TO1) and E(TO1) are components of the soft TO1 mode.

FIG. 8. (Color online) The same as in Fig. 7 for temperatures above 600 K. E(TO1)+CM denotes a joint contribution of the E component and relaxational central mode (CM) at the lowest frequencies.

FIG. 9. (Color online) Complex dielectric response of the FGC in the THz and IR range (solid lines) obtained by fitting of the IR reflectivity spectra together with the THz experimental data (symbols) at temperatures below 450 K.

FIG. 10. (Color online) The same as in Fig. 9 for temperatures above 500 K.

FIG. 11. (Color online) Components of the soft TO1 phonon mode. Temperature dependences of the $A_1$(TO1) and E(TO1) frequencies in the CGC (solid lines and symbols) and FGC (dash lines and open symbols) obtained from the multi-oscillator fits of the IR reflectivity spectra together with the THz data.

FIG. 12. (Color online) Soft and central modes in the CGC (solid lines and symbols) and FGC (dash lines and open symbols). Characteristic frequencies of the relaxational central mode components (R1, R2, R) and soft E(TO1) mode correspond to the maxima in the dielectric loss spectra. Symbols denote the experimental points, lines R1, R2 and R are fits of their temperature dependences (see details in the text).

FIG. 13. (Color online) Temperature dependence of the SHG signal in the CGC (solid lines) and FGC (dash lines).

TABLE I. Parameters of polar phonon modes in FGC. All the parameters are in cm$^{-1}$.

| 5 K | | | | 800 K | | | |
|---|---|---|---|---|---|---|---|
| $\omega_{TO}$ | $\gamma_{TO}$ | $\omega_{LO}$ | $\gamma_{LO}$ | $\omega_{TO}$ | $\gamma_{TO}$ | $\omega_{LO}$ | $\gamma_{LO}$ |
| 32.9 | 44.8 | 46 | 35.8 | 30.8 | 77.4 | 49.5 | 224.8 |
| 53 | 15.4 | 54 | 15.9 | | | | |
| 76.9 | 30.9 | 76.9 | 15.5 | | | | |
| 80.6 | 20.7 | 103 | 26.9 | 50.3 | 96.8 | 117 | 37.5 |
| 105.8 | 29.4 | 125 | 26.3 | | | | |
| 152 | 103.9 | 175.8 | 66.3 | | | | |
| 211.8 | 69.8 | 291 | 212 | 209.5 | 143.8 | 286.3 | 262.8 |
| 292.9 | 87.5 | 334.9 | 96.2 | | | | |
| 350.6 | 83.8 | 384.3 | 43.4 | 319.8 | 227.1 | 361.4 | 84.5 |
| 386.7 | 35 | 413.1 | 28.8 | 371.8 | 95.9 | 413.5 | 33.9 |
| 451.5 | 93.4 | 458.8 | 29.9 | 451.3 | 91.6 | 451.5 | 50.1 |
| 524.4 | 110.4 | 616 | 200.1 | | | | |
| 620.1 | 93.9 | 714.3 | 52.3 | 492.7 | 156.6 | 703.3 | 58.6 |

Table II.

Fit parameters of the temperature dependences of the relaxation frequencies in CGS and FGS.

| Relaxation frequency | Temperature range (Equation) | Critical slowing down | | Vogel-Fulcher law | | Arrhenius law | |
|---|---|---|---|---|---|---|---|
| | | $A_1$, $A_2$ [GHz/K] | $T_1$, $T_2$ [K] | $U_1$, $U_2$ [K] | $T_{VF1}$, $T_{VF2}$ [K] | $\nu_0$ [THz] | $E_1$, $E_2$ [K] |
| $\nu_{R1}$ CGC | below $T_C$ | 25 | 478 | | | | 1500 |
| | above $T_C$ | 9 | 400 | | | | 1500 |
| $\nu_{R2}$ CGC | below $T_C$ (6a) | 0.3 | 447 | 700 | 190 | | |
| | above $T_C$ (6b) | 6 | 463 | 700 | 210 | | |
| $\nu_R$ FGC | whole range | | | | | 14 | 2600 |

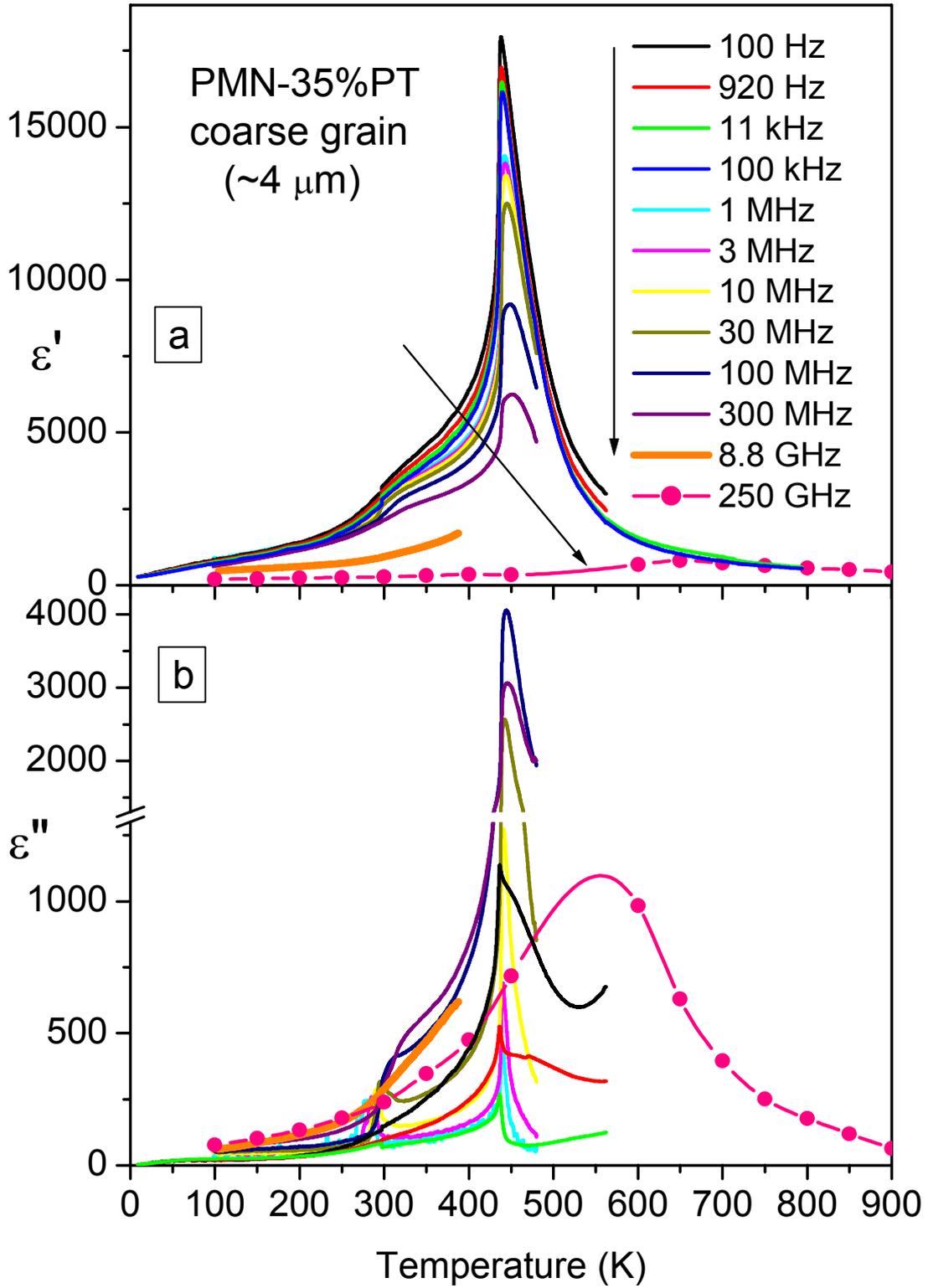

FIG. 1. (Color online) Temperature dependences of (a) dielectric permittivity $\varepsilon'$ and (b) loss $\varepsilon''$ of the CGC at frequencies between 100 Hz and 250 GHz.

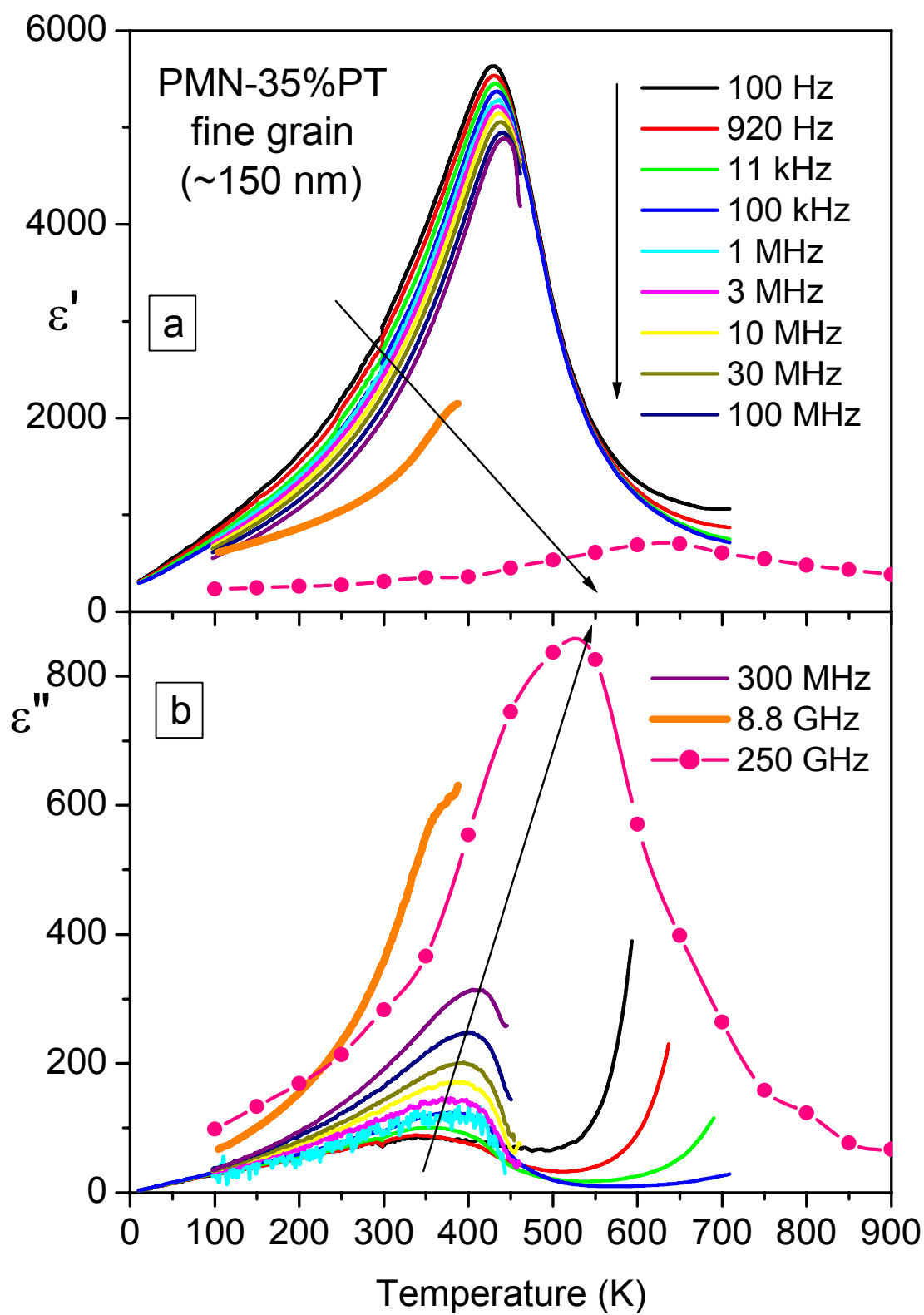

FIG. 2. (Color online) Temperature dependences of (a) dielectric permittivity $\varepsilon'$ and (b) loss $\varepsilon''$ of the FGC at frequencies between 100 Hz and 250 GHz.

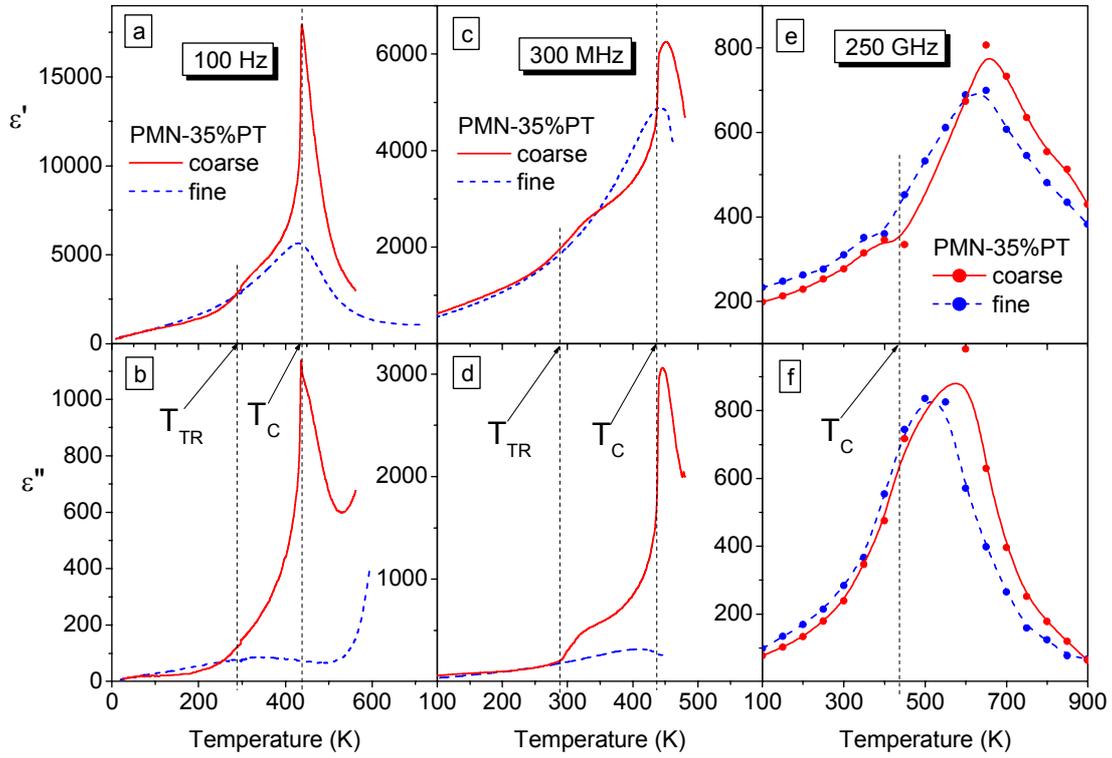

FIG. 3. (Color online) Temperature dependences of the dielectric permittivity $\varepsilon'$ (a, c, e) and loss $\varepsilon''$ (b, d, f) of the CGC (solid lines) and FGC (dash lines) at 100 Hz (a, b), 300 MHz (c, d) and 250 GHz (e, f).

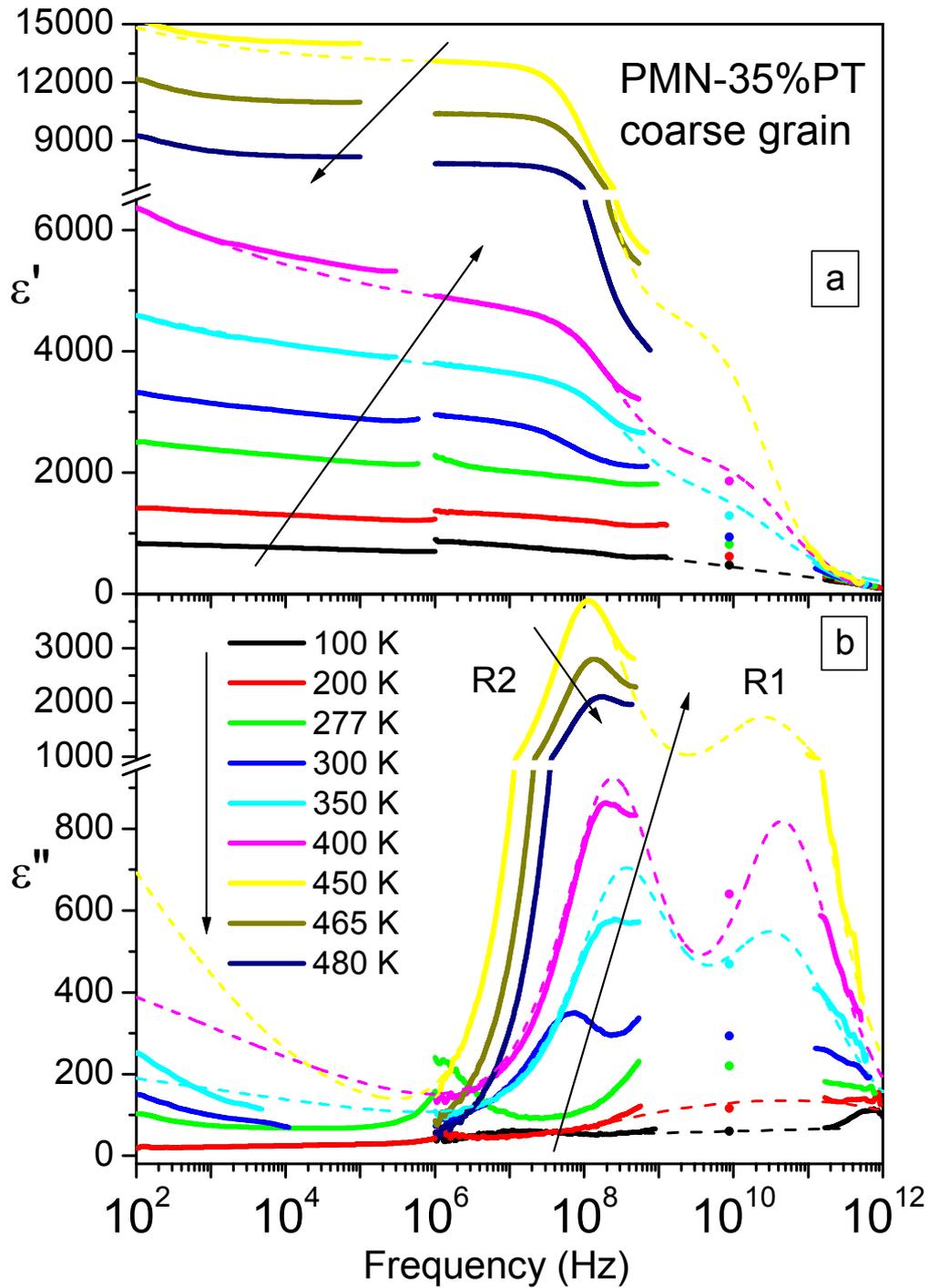

FIG. 4. (Color online) Frequency dependences of (a) dielectric permittivity $\varepsilon'$ and (b) loss $\varepsilon''$ of the CGC at temperatures between 100 K and 480 K. Dash lines at $T \geq 300$ K are multi-component Cole-Cole fits. R1 and R2 denote loss maxima of the two relaxational contributions (see the text).

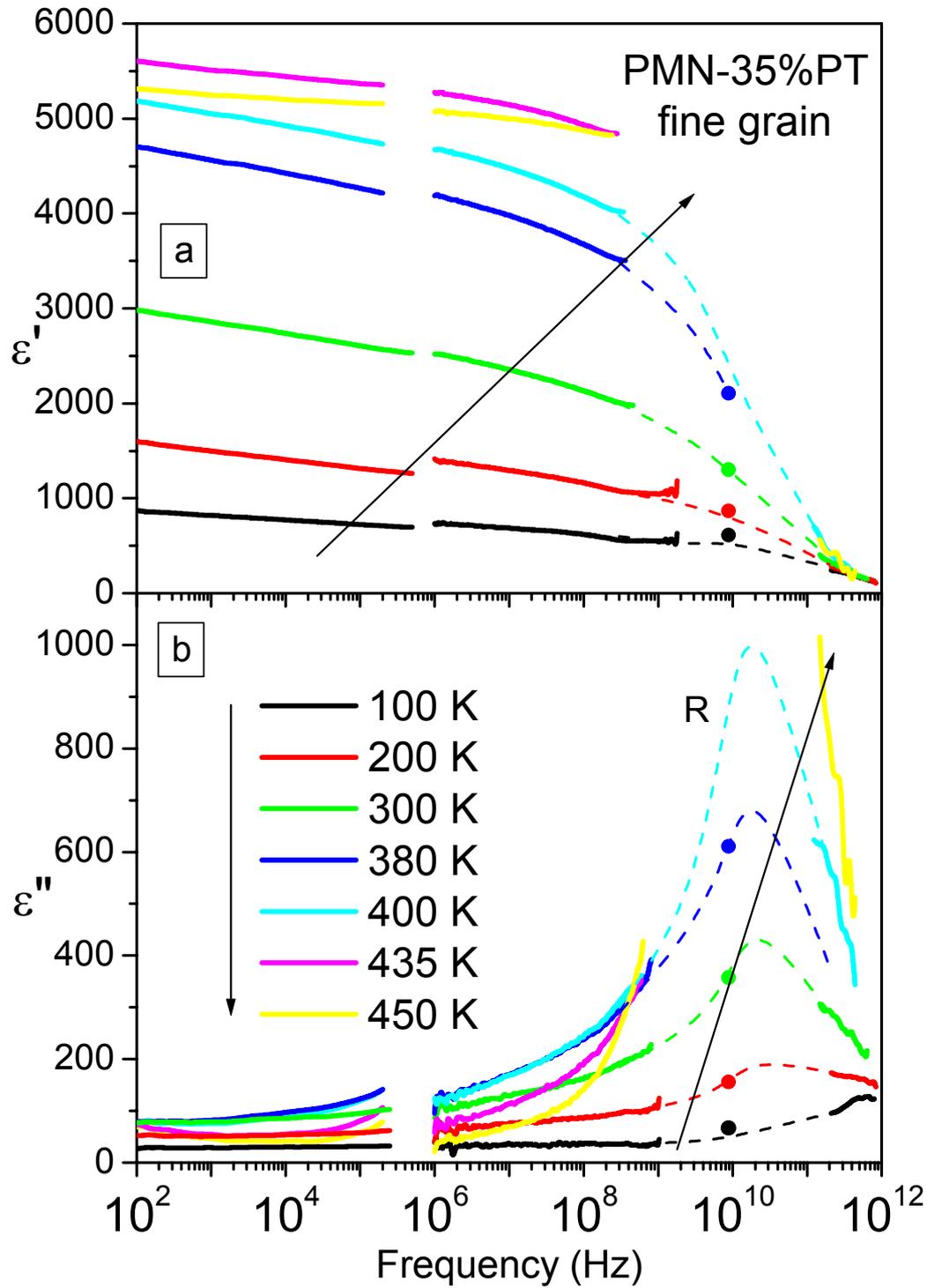

FIG. 5. (Color online) Frequency dependences of (a) dielectric permittivity $\varepsilon'$ and (b) loss $\varepsilon''$ of the FGC at temperatures between 100 K and 450 K. R denotes the asymmetric loss maximum by overlapping two relaxational contributions (see the text).

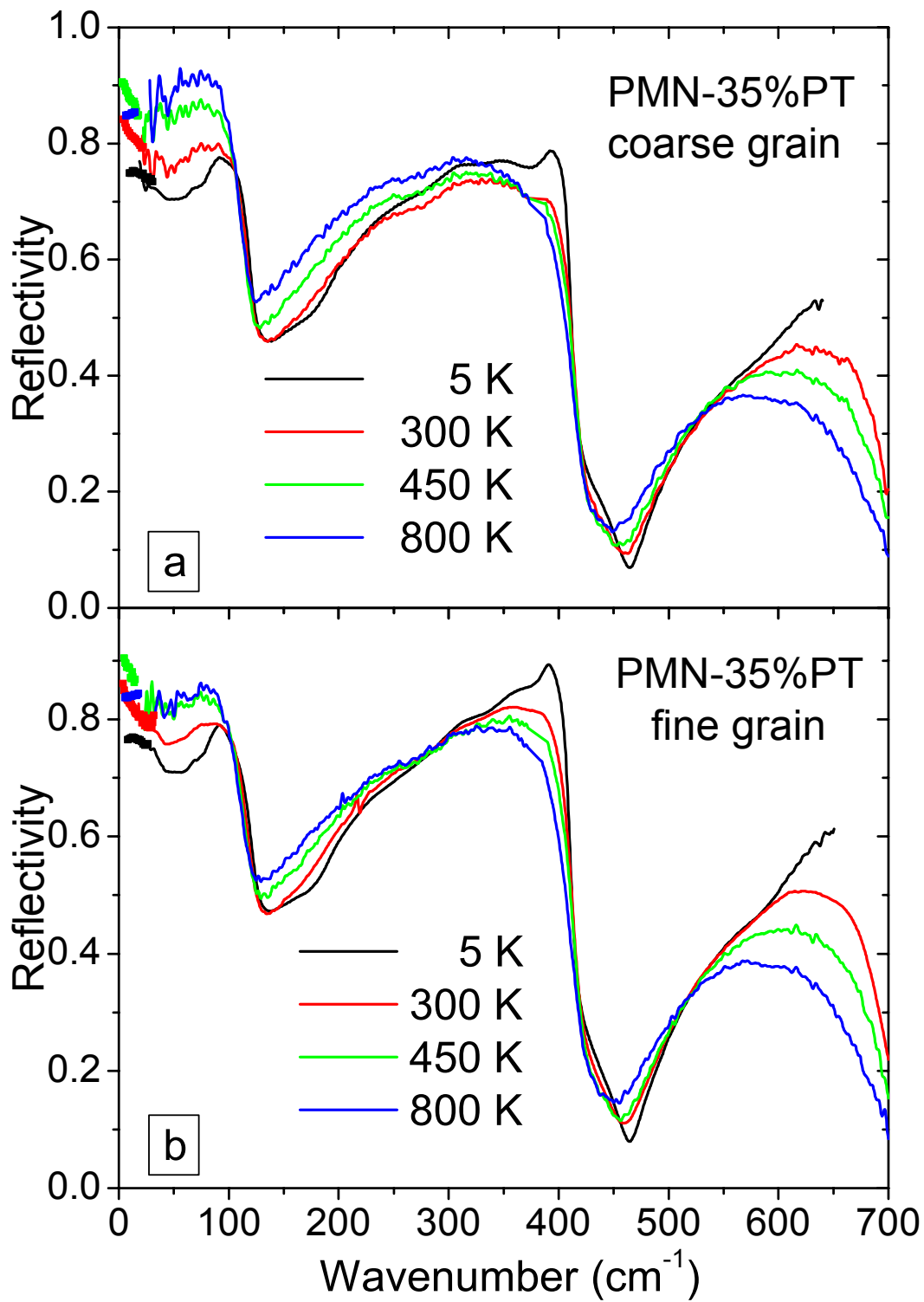

FIG. 6. (Color online) IR and THz reflectivity spectra of (a) CGC and (b) FGC at temperatures between 5 K and 800 K. THz reflectivities were calculated from the THz dielectric response.

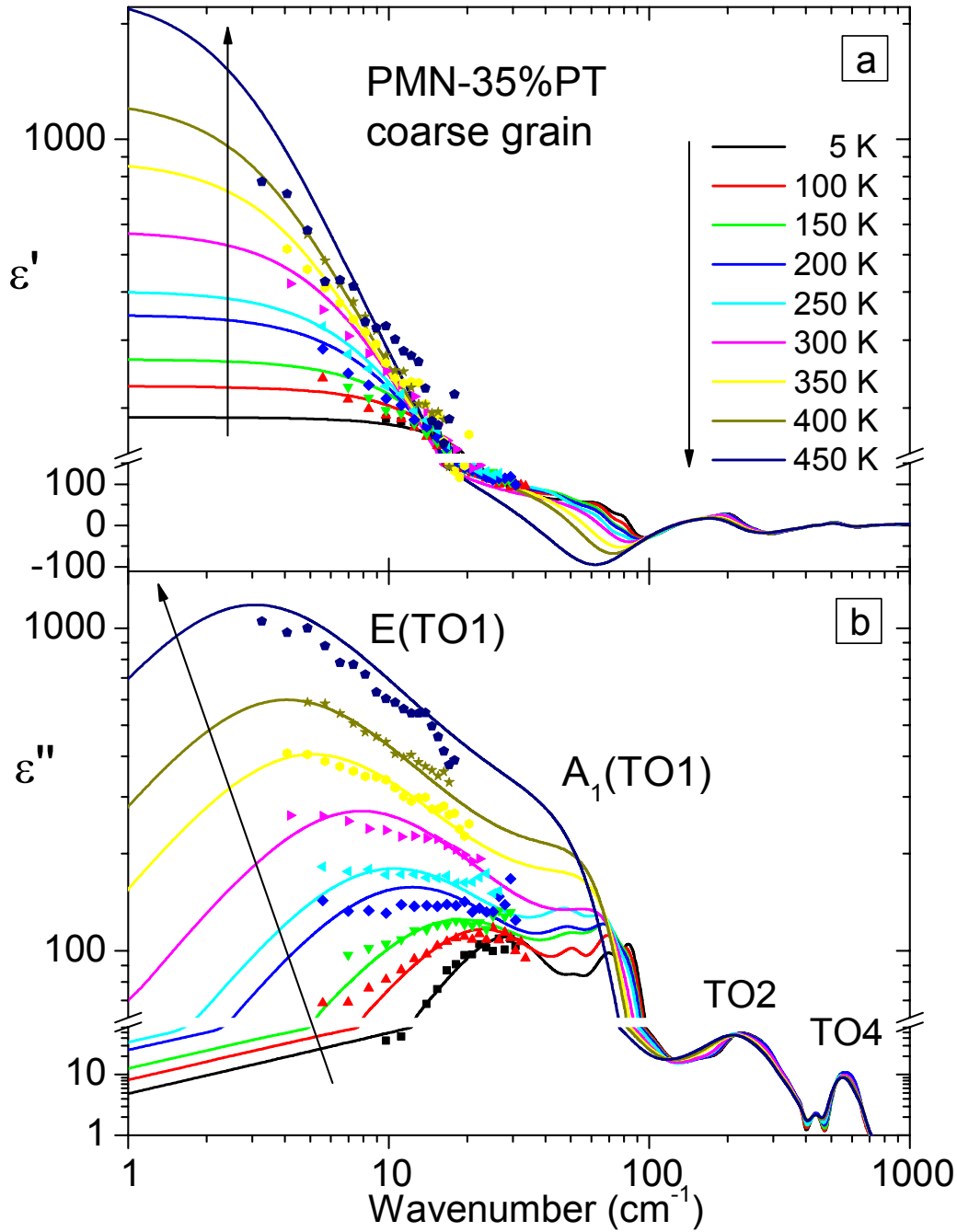

FIG. 7. (Color online) Complex dielectric response of the CGC in the THz and IR range (solid lines) obtained by fitting of the IR reflectivity spectra together with THz experimental data (symbols) at temperatures below 450 K. TO1, TO2 and TO4 denote the loss maxima attributed to the 3 IR-active transverse optical phonons. $A_1$(TO1) and $E$(TO1) are components of the soft TO1 mode.

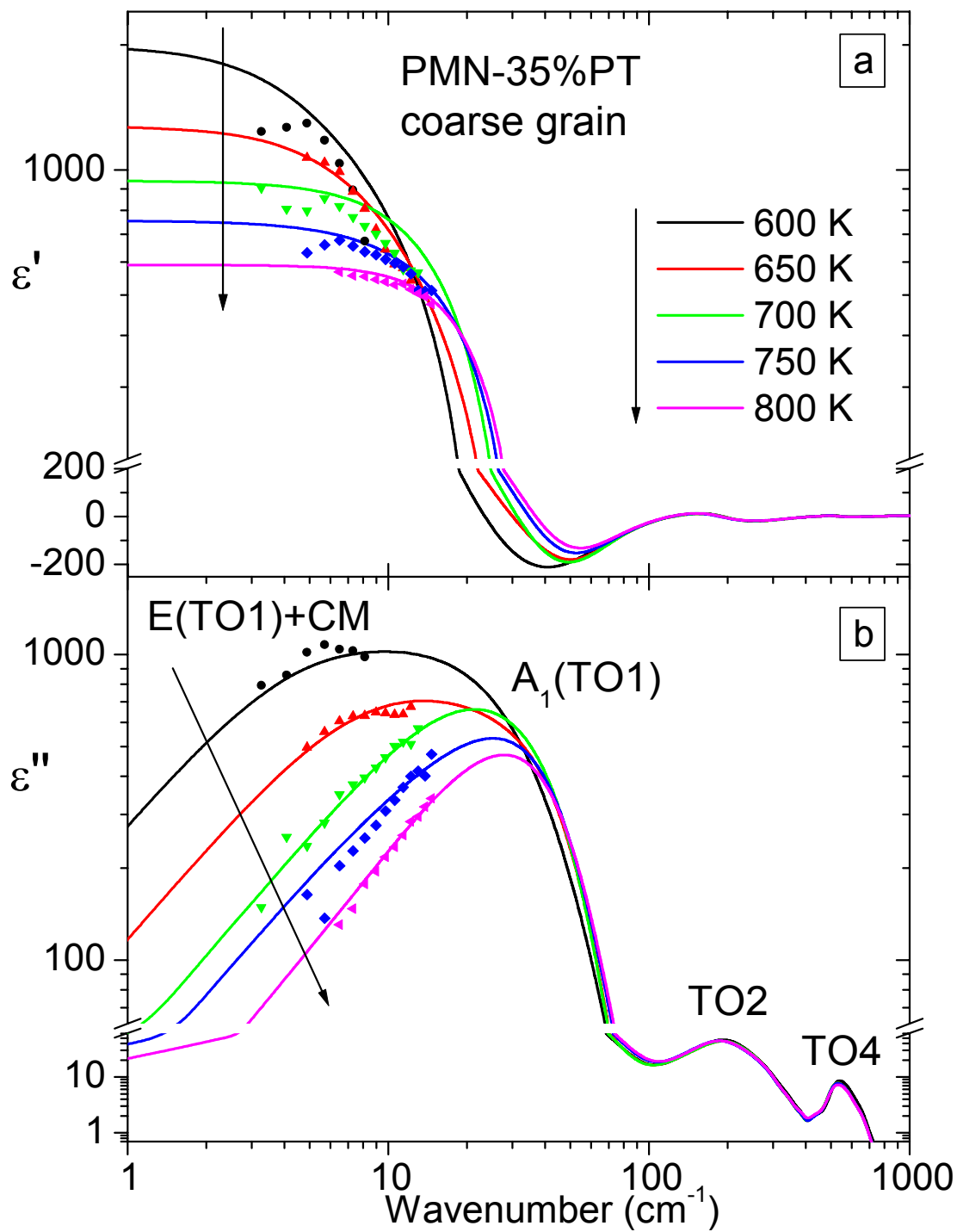

FIG. 8. (Color online) The same as in Fig. 7 for temperatures above 600 K. E(TO1)+CM denotes a joint contribution of the E component and relaxational central mode (CM) at the lowest frequencies.

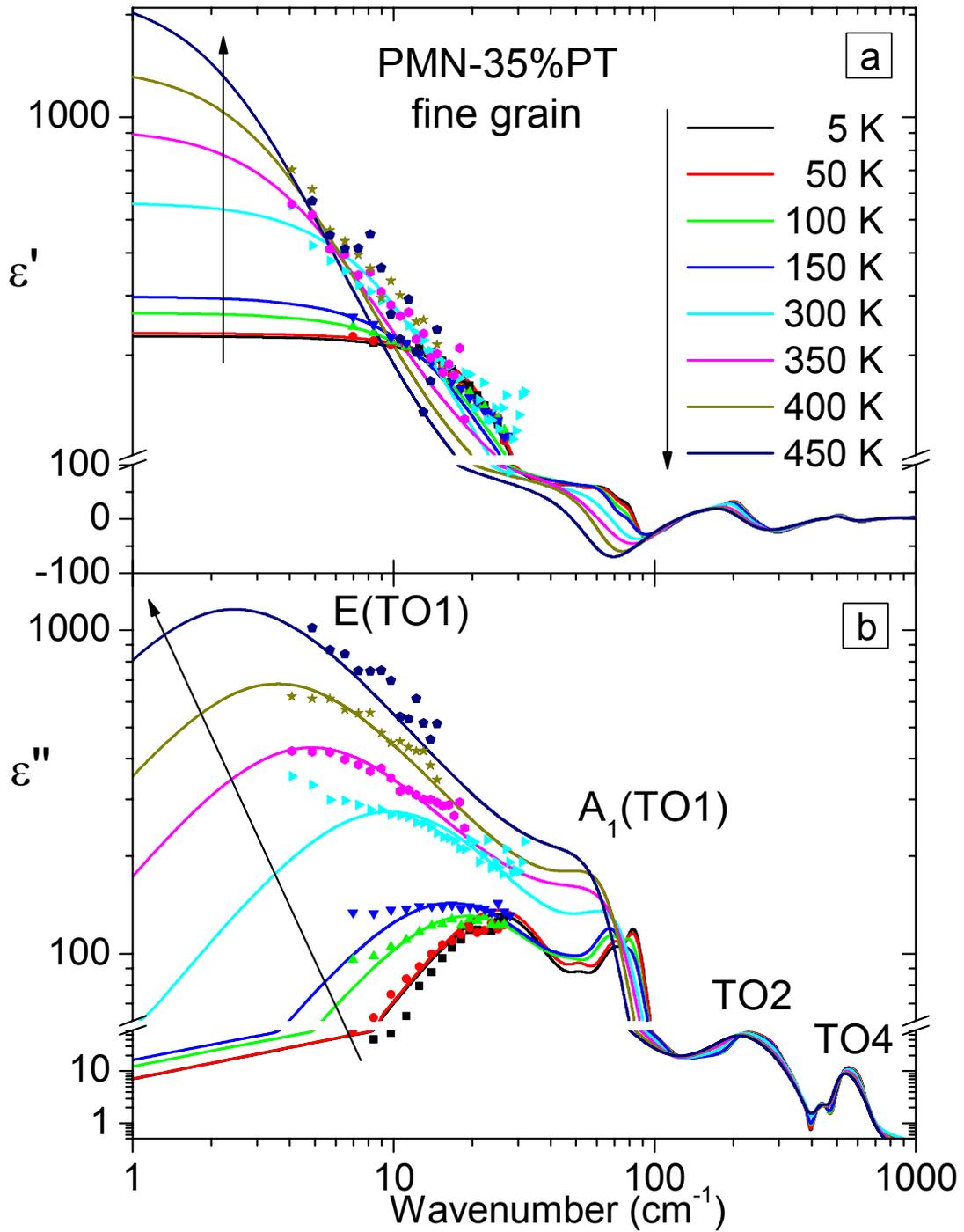

FIG. 9. (Color online) Complex dielectric response of the FGC in the THz and IR range (solid lines) obtained by fitting of the IR reflectivity spectra together with the THz experimental data (symbols) at temperatures below 450 K.

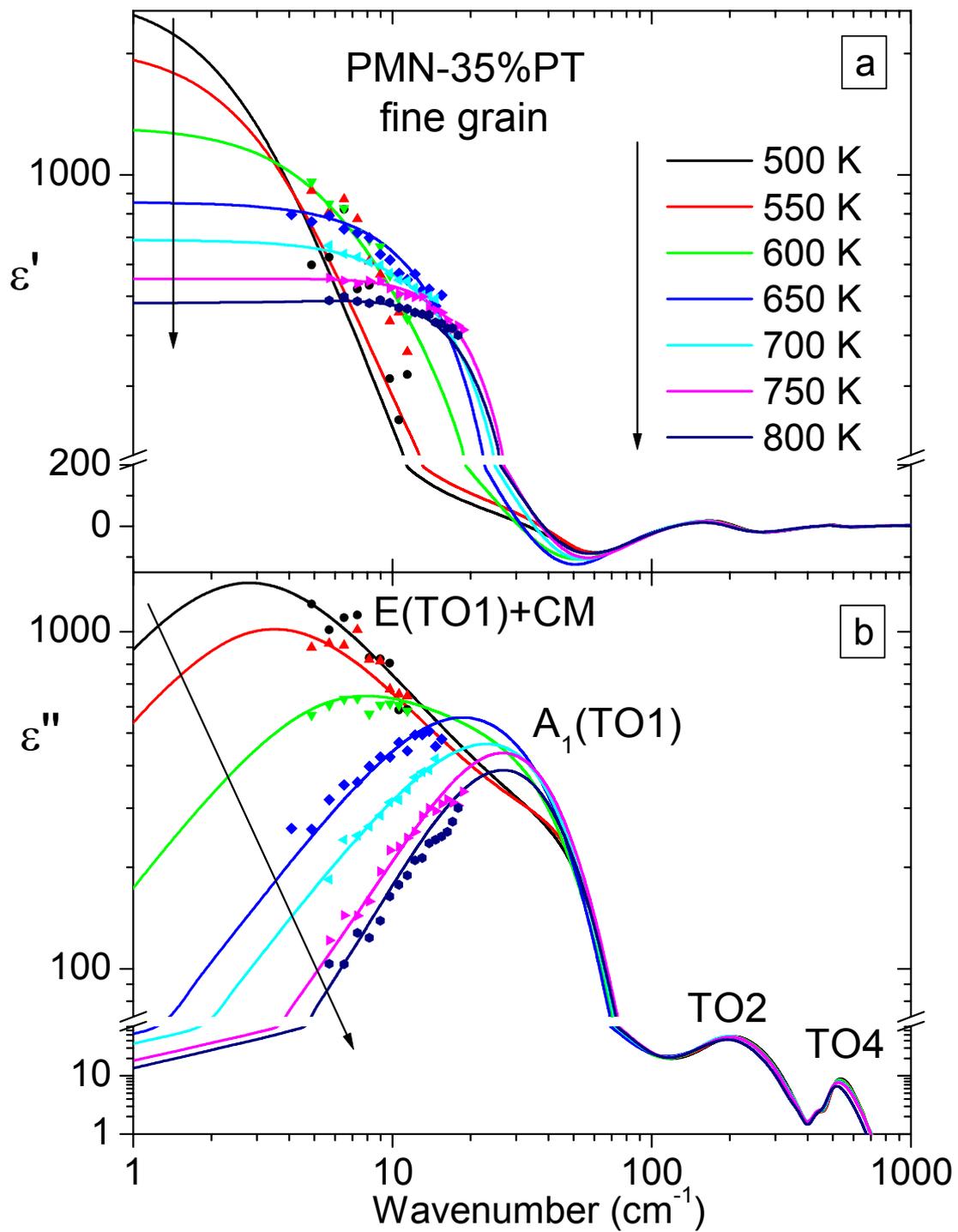

FIG. 10. (Color online) The same as in Fig. 9 for temperatures above 500 K.

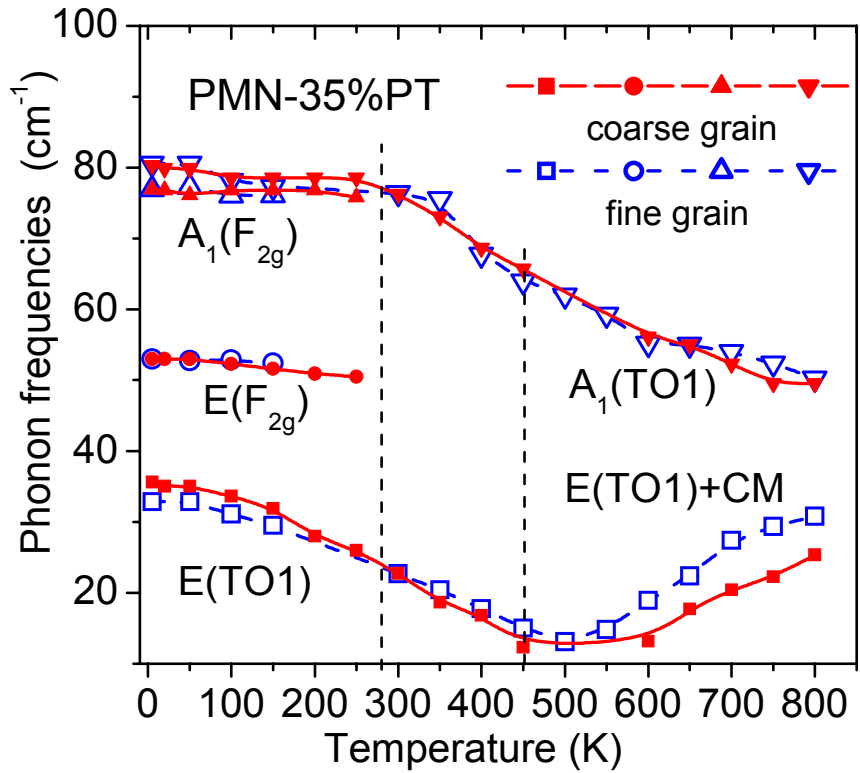

FIG. 11. (Color online) Components of the soft TO1 phonon mode. Temperature dependences of the $A_1$(TO1) and E(TO1) frequencies in the CGC (solid lines and symbols) and FGC (dash lines and open symbols) obtained from the multi-oscillator fits of the IR reflectivity spectra together with the THz data.

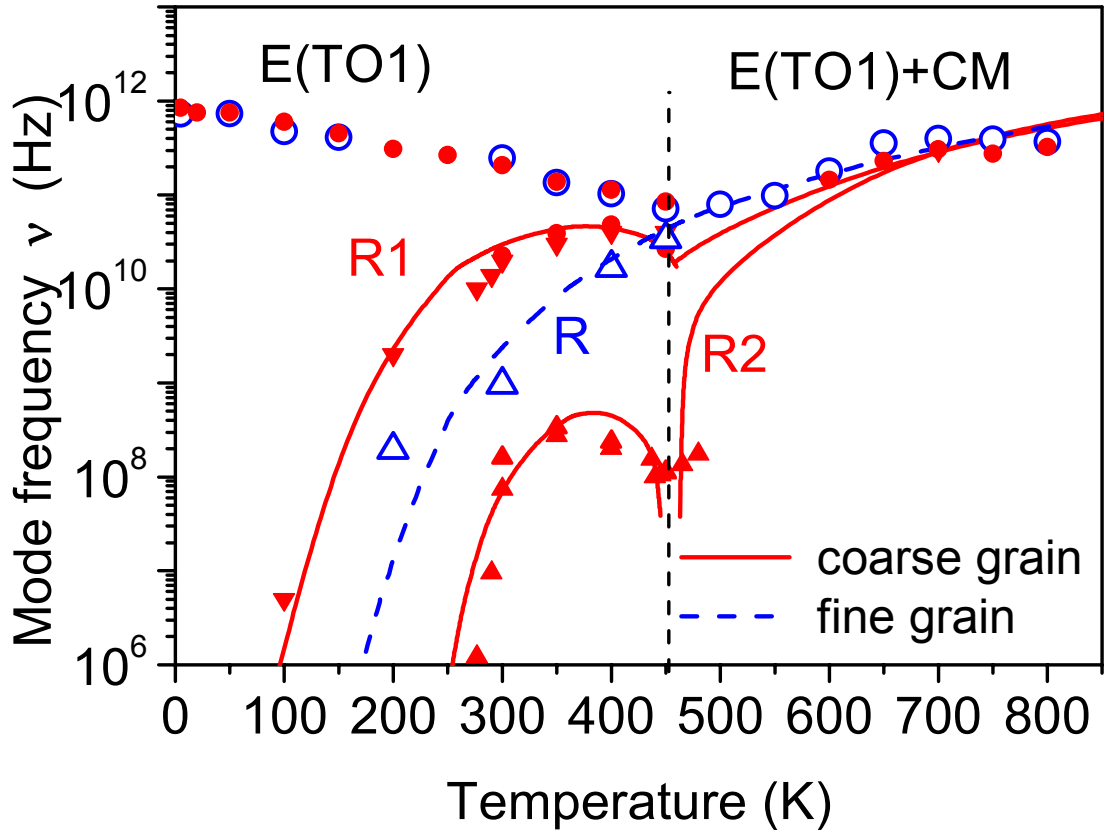

FIG. 12. (Color online) Soft and central modes in the CGC (solid lines and symbols) and FGC (dash lines and open symbols). Characteristic frequencies of the relaxational central mode components (R1, R2, R) and soft E(TO1) mode correspond to the maxima in the dielectric loss spectra. Symbols denote the experimental points, lines R1, R2 and R are fits of their temperature dependences (see details in the text).

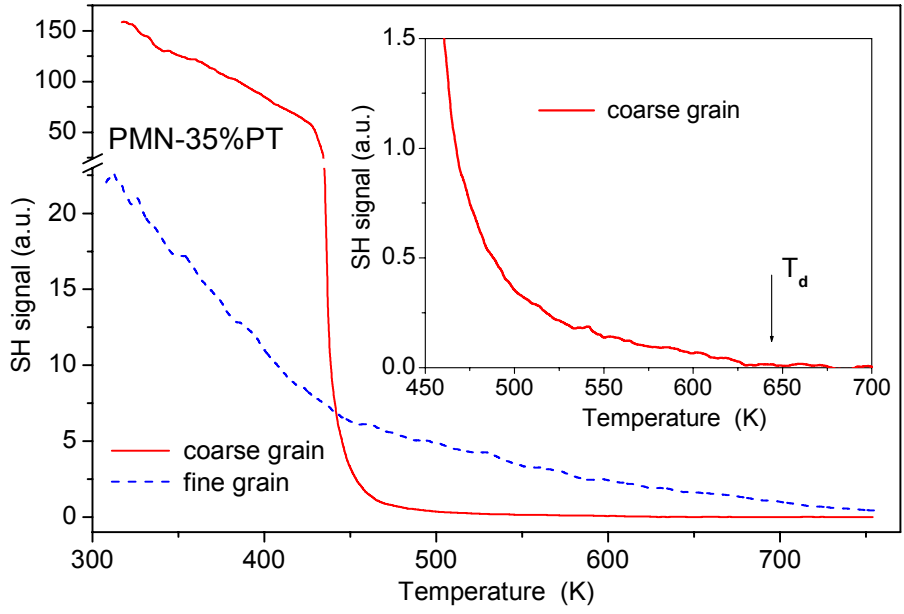

FIG. 13. (Color online) Temperature dependence of the second harmonic generation signal in the CGC (solid lines) and FGC (dash lines).